\documentstyle[12pt]{article}
\begin{document}
\def\RN{Reisner-Nordstr\"om }
\def\Sch{Schwarzchild }
\def\HH{Hartle-Hawking }
\def\BF{Balbinot-Fabbri }
\def\PoL{Polyakov-Liouville }
\def\be{\begin{equation}} \def\bea{\begin{eqnarray}}
\def\ee{\end{equation}}\def\eea{\end{eqnarray}}
\def\ncr{\nonumber\\ }
\def\nn{{g_{00}}} \def\nj{{g_{01}}} \def\jj{{g_{11}}}
\def\nnp{{g_{00}^\prime}} \def\njp{{g_{01}^\prime}} \def\jjp{{g_{11}^\prime}}
\def\rp{{r^\prime}} \def\psip{{\psi ^\prime}} \def\chip{{\chi ^\prime}}
\def\raz{{g_{01}\over g_{11}}}
\def\kr{(\dot r-\raz\rp )}
\def\kpsi{(\dot\psi -\raz\psip )}
\def\kchi{(\dot\chi -\raz\chip )}
\def\kg{(\raz\jjp +\dot\jj -2\njp )}
\def\sg{\sqrt{-g}}
\def\H{{\cal H}}
\def\MBVR{Maja Buri\'c
\footnote{E-mail: majab@rudjer.ff.bg.ac.yu} and
Voja Radovanovi\'c\footnote{E-mail: rvoja@rudjer.ff.bg.ac.yu}\\
{\it Faculty of Physics, P.O. Box 368, 11001 Belgrade, Yugoslavia}}


 
\let\und=\b                     
\let\ced=\c                     
\let\du=\d                      
\let\um=\H                      
\let\sll=\l                     
\let\Sll=\L                     
\let\slo=\o                     
\let\Slo=\O                     
\let\tie=\t                     
\let\br=\u                      

 
\def\a{\alpha}
\def\b{\beta}
\def\c{\chi}
\def\d{\delta}
\def\e{\epsilon}                
\def\f{\phi}                    
\def\g{\gamma}
\def\h{\eta}
\def\i{\iota}
\def\j{\psi}
\def\k{\kappa}
\def\l{\lambda}
\def\m{\mu}
\def\n{\nu}
\def\o{\omega}
\def\p{\pi}                     
\def\q{\theta}                  
\def\r{\rho}                    
\def\s{\sigma}                  
\def\t{\tau}
\def\u{\upsilon}
\def\x{\xi}
\def\z{\zeta}
\def\D{\Delta}
\def\F{\Phi}
\def\G{\Gamma}
\def\J{\Psi}
\def\L{\Lambda}
\def\O{\Omega}
\def\P{\Pi}
\def\Q{\Theta}
\def\S{\Sigma}
\def\U{\Upsilon}
\def\X{\Xi}
 
 
\def\ca{{\cal A}}
\def\cb{{\cal B}}
\def\cc{{\cal C}}
\def\cd{{\cal D}}
\def\ce{{\cal E}}
\def\cf{{\cal F}}
\def\cg{{\cal G}}
\def\ch{{\cal H}}
\def\ci{{\cal I}}
\def\cj{{\cal J}}
\def\ck{{\cal K}}
\def\cl{{\cal L}}
\def\cm{{\cal M}}
\def\cn{{\cal N}}
\def\co{{\cal O}}
\def\cp{{\cal P}}
\def\cq{{\cal Q}}
\def\car{{\cal R}}
\def\cs{{\cal S}}
\def\ct{{\cal T}}
\def\cu{{\cal U}}
\def\cv{{\cal V}}
\def\cw{{\cal W}}
\def\cx{{\cal X}}
\def\cy{{\cal Y}}
\def\cz{{\cal Z}}

 
\def\bo{{\raise.05ex\hbox{\large$\Box$}\:}}             
\def\cbo{{\,\raise-.15ex\Sc [\,}}                       
\def\pa{\partial}                                       
\def\de{\nabla}                                         
\def\dell{\bigtriangledown}                             
\def\su{\sum}                                           
\def\pr{\prod}                                          
\def\iff{\leftrightarrow}                               
\def\conj{{\hbox{\large *}}}                            
\def\ltap{\raisebox{-.4ex}{\rlap{$\sim$}} \raisebox{.4ex}{$<$}}   
\def\gtap{\raisebox{-.4ex}{\rlap{$\sim$}} \raisebox{.4ex}{$>$}}   
\def\TH{{\raise.2ex\hbox{$\displaystyle \bigodot$}\mskip-4.7mu \llap H \;}}
\def\face{\hbox{\normalsize$\;\;\:{\raise.9ex\hbox{\oo n}\mskip-13mu \llap
        {${\buildrel{\hbox{\frtnrm ..}}\over\smile}$}}\:$}}     
\def\Face{{\raise.2ex\hbox{$\displaystyle \bigodot$}\mskip-2.2mu \llap {$\ddot
        \smile$}}}                                      
\def\dg{\sp\dagger}                                     
\def\ddg{\sp\ddagger}                                   
\def\Lhat{{\bf\rlap{\kern-.09em$\hat{\phantom L}$}L}}
\def\Lcheck{{\bf\rlap{\kern-.09em$\check{\phantom L}$}L}}
 
 
\def\sp#1{{}^{#1}}                              
\def\sb#1{{}_{#1}}                              
\def\oldsl#1{\rlap/#1}                          
\def\sl#1{\rlap{\hbox{$\mskip 1 mu /$}}#1}      
\def\Sl#1{\rlap{\hbox{$\mskip 3 mu /$}}#1}      
\def\SL#1{\rlap{\hbox{$\mskip 4.5 mu /$}}#1}    
\def\Tilde#1{\widetilde{#1}}                    
\def\Hat#1{\widehat{#1}}                        
\def\Bar#1{\overline{#1}}                       
\def\bra#1{\Big\langle #1\Big|}                 
\def\ket#1{\Big| #1\Big\rangle}                 
\def\VEV#1{\Big\langle #1\Big\rangle}           
\def\brak#1#2{\Big\langle #1\Big|#2\Big\rangle}         
\def\abs#1{\Big| #1\Big|}                       
\def\sbra#1{\left\langle #1\right|}             
\def\sket#1{\left| #1\right\rangle}             
\def\svev#1{\left\langle #1\right\rangle}       
\def\sabs#1{\left| #1\right|}                   

\def\leftrightarrowfill{$\mathsurround=0pt \mathord\leftarrow \mkern-6mu
        \cleaders\hbox{$\mkern-2mu \mathord- \mkern-2mu$}\hfill
        \mkern-6mu \mathord\rightarrow$}
\def\dvec#1{\vbox{\ialign{##\crcr
        \leftrightarrowfill\crcr\noalign{\kern-1pt\nointerlineskip}
        $\hfil\displaystyle{#1}\hfil$\crcr}}}           
\def\dt#1{{\buildrel {\hbox{\LARGE .}} \over {#1}}}     
\def\dtt#1{{\buildrel \bullet \over {#1}}}              
\def\ddt#1{{\buildrel {\hbox{\LARGE .\kern-2pt.}} \over {#1}}}
\def\der#1{{\pa \over \pa {#1}}}                
\def\fder#1{{\d \over \d {#1}}}                 
 
 
\def\frac#1#2{{\textstyle{#1\over\vphantom2\smash{\raise.20ex
        \hbox{$\scriptstyle{#2}$}}}}}                   
\def\ha{\frac12}                                        
\def\sfrac#1#2{{\vphantom1\smash{\lower.5ex\hbox{\small$#1$}}\over
        \vphantom1\smash{\raise.4ex\hbox{\small$#2$}}}} 
\def\bfrac#1#2{{\vphantom1\smash{\lower.5ex\hbox{$#1$}}\over
        \vphantom1\smash{\raise.3ex\hbox{$#2$}}}}       
\def\afrac#1#2{{\vphantom1\smash{\lower.5ex\hbox{$#1$}}\over#2}}    
\def\tder#1#2{{d #1 \over d #2 }}                 
\def\partder#1#2{{\partial #1\over\partial #2}}   
\def\brkt#1#2{{\left\langle #1 | #2 \right\rangle}} 
\def\secder#1#2#3{{\partial~2 #1\over\partial #2 \partial #3}}  
\def\on#1#2{\mathop{\null#2}\limits~{#1}}       
\def\On#1#2{{\buildrel{#1}\over{#2}}}           
\def\under#1#2{\mathop{\null#2}\limits_{#1}}    
\def\bvec#1{\on\leftarrow{#1}}                  
\def\oover#1{\on\circ{#1}}                              
 
 
\def\boxes#1{
        \newcount\num
        \num=1
        \newdimen\downsy
        \downsy=-1.64ex
        \mskip-7.8mu
        \bo
        \loop
        \ifnum\num<#1
        \llap{\raise\num\downsy\hbox{$\bo$}}
        \advance\num by1
        \repeat}
\def\boxup#1#2{\newcount\numup
        \numup=#1
        \advance\numup by-1
        \newdimen\upsy
        \upsy=.82ex
        \mskip7.8mu
        \raise\numup\upsy\hbox{$#2$}}
 
 
\newskip\humongous \humongous=0pt plus 1000pt minus 1000pt
\def\caja{\mathsurround=0pt}
\def\eqalign#1{\,\vcenter{\openup2\jot \caja
        \ialign{\strut \hfil$\displaystyle{##}$&$
        \displaystyle{{}##}$\hfil\crcr#1\crcr}}\,}
\newif\ifdtup
\def\panorama{\global\dtuptrue \openup2\jot \caja
        \everycr{\noalign{\ifdtup \global\dtupfalse
        \vskip-\lineskiplimit \vskip\normallineskiplimit
        \else \penalty\interdisplaylinepenalty \fi}}}
\def\li#1{\panorama \tabskip=\humongous                         
        \halign to\displaywidth{\hfil$\displaystyle{##}$
        \tabskip=0pt&$\displaystyle{{}##}$\hfil
        \tabskip=\humongous&\llap{$##$}\tabskip=0pt
        \crcr#1\crcr}}
\def\eqalignnotwo#1{\panorama \tabskip=\humongous
        \halign to\displaywidth{\hfil$\displaystyle{##}$
        \tabskip=0pt&$\displaystyle{{}##}$
        \tabskip=0pt&$\displaystyle{{}##}$\hfil
        \tabskip=\humongous&\llap{$##$}\tabskip=0pt
        \crcr#1\crcr}}
 
 
\def\phil{@{\extracolsep{\fill}}}
\def\unphil{@{\extracolsep{\tabcolsep}}}
 

\def\CMP{Commun. Math. Phys.}
\def\NP{Nucl. Phys. B\,}
\def\PL{Phys. Lett. B\,}
\def\PR{Phys. Rev. Lett.}
\def\PRD{Phys. Rev. D\,}
\def\CQG{Class. Quant. Grav.}
\def\IJMP{Int. J. Mod. Phys.}
\def\MPL{Mod. Phys. Lett.}

\def\ref#1{$\sp{#1]}$}
\def\Ref#1{$\sp{#1)}$}
 
 
\topmargin=.17in                        
\headheight=0in                         
\headsep=0in                    
\textheight=9in                         
\footheight=3ex                         
\footskip=4ex           
\textwidth=6in                          
\hsize=6in                              
\parindent=21pt                         
\parskip=\medskipamount                 
\lineskip=0pt                           
\abovedisplayskip=1em plus.3em minus.5em        
\belowdisplayskip=1em plus.3em minus.5em        
\abovedisplayshortskip=.5em plus.2em minus.4em  
\belowdisplayshortskip=.5em plus.2em minus.4em  
\def\baselinestretch{1.2}       
\thicklines                         
\oddsidemargin=.25in \evensidemargin=.25in      
\marginparwidth=.85in                           
 
 
\def\title#1#2#3#4{
        {\hbox to\hsize{#4 \hfill  #3}}\par
        \begin{center}\vskip.5in minus.1in {\Large\bf #1}\\[.5in minus.2in]{#2}
        \vskip1.4in minus1.2in {\bf ABSTRACT}\\[.1in]\end{center}
        \begin{quotation}\par}
\def\author#1#2{#1\\[.1in]{\it #2}\\[.1in]}

\def\AMIC{Aleksandar Mikovic\'c
\\[.1in]{\it Blackett Laboratory, Imperial College, Prince Consort Road, London
SW7 2BZ, UK}\\[.1in]}

\def\AMICIF{Aleksandar Mikovi\'c\,
\footnote{Work supported by MNTRS and Royal Society}
\\[.1in] {\it Blackett Laboratory, Imperial College, Prince Consort
Road, London SW7 2BZ, UK}\\[.1in]
and \\[.1 in]
{\it Institute of Physics, P.O. Box 57, 11001 Belgrade, Yugoslavia}
\footnote{Permanent address}\\ {\it E-mail:\, mikovic@castor.phy.bg.ac.yu}}

\def\AMSISSA{Aleksandar Mikovi\'c\,
\footnote{E-mail address: mikovic@castor.phy.bg.ac.yu}
\\[.1in] {\it SISSA-International School for Advanced Studies\\
Via Beirut 2-4, Trieste 34100, Italy}\\[.1in]
and \\[.1 in]
{\it Institute of Physics, P.O. Box 57, 11001 Belgrade, Yugoslavia}
\footnote{Permanent address}}

\def\AM{Aleksandar Mikovi\'c 
\footnote{E-mail address: mikovic@castor.phy.bg.ac.yu}
\\[.1in] {\it Institute of Physics, P.O.Box 57, Belgrade 11001, Yugoslavia}
\\[.1in]}

\def\AMsazda{Aleksandar Mikovi\'c 
\footnote{E-mail address: mikovic@castor.phy.bg.ac.yu}
and Branislav Sazdovi\'c \footnote{E-mail: sazdovic@castor.phy.bg.ac.yu}
\footnote{Work supported by MNTRS}
\\[.1in] {\it Institute of Physics, P.O.Box 57, Belgrade 11001, Yugoslavia}
\\[.1in]}

\def\AMVR{Aleksandar Mikovi\'c\,
\footnote{E-mail address: mikovic@castor.phy.bg.ac.yu}
\\[.1in] 
{\it Institute of Physics, P.O. Box 57, 11001 Belgrade, Yugoslavia}
\\[.2in]
Voja Radovanovi\'c \\[.1 in]
{\it Faculty of Physics, P.O. Box 550, 11001 Belgrade, Yugoslavia}}

\def\AMCVR{Aleksandar Mikovi\'c
\footnote{Permanent address: Institute of Physics, P.O. Box 57, 11001 
Belgrade, Yugoslavia}\footnote{E-mail: mikovic@fy.chalmers.se, 
mikovic@castor.phy.bg.ac.yu}
\\
{\it Institute of Theoretical Physics, Chalmers University of Technology,
S-412 96 Goteborg, Sweden}\\[.1in]
and
\\[.1in]
Voja Radovanovi\'c
\footnote{E-mail: rvoja@rudjer.ff.bg.ac.yu} \\
{\it Faculty of Physics, P.O. Box 550, 11001 Belgrade, Yugoslavia}}

\def\AMVVR{Aleksandar Mikovi\'c
\footnote{On leave from Institute of Physics, P.O. Box 57, 11001 
Belgrade, Yugoslavia}
\footnote{Supported by Comissi\'on Interministerial de Ciencia y Tecnologia}
\footnote{E-mail: mikovic@lie1.ific.uv.es}
\\
{\it Departamento de Fisica Te\'orica and IFIC, Centro Mixto Universidad
de Valencia-CSIC, Facultad de Fisica, Burjassot-46100, Valencia, Spain}
\\[.1in]
Voja Radovanovi\'c
\footnote{E-mail: rvoja@rudjer.ff.bg.ac.yu} \\
{\it Faculty of Physics, P.O. Box 368, 11001 Belgrade, Yugoslavia}}

\def\MBAMVVR{Maja Buri\'c
\footnote{E-mail: majab@rudjer.ff.bg.ac.yu}\\
{\it Faculty of Physics, P.O. Box 368, 11001 Belgrade, Yugoslavia}
\\[.1in]
Aleksandar Mikovi\'c
\footnote{On leave from Institute of Physics, P.O. Box 57, 11001 
Belgrade, Yugoslavia}
\footnote{Supported by Comissi\'on Interministerial de Ciencia y Tecnologia}
\footnote{E-mail: mikovic@lie1.ific.uv.es}
\\
{\it Departamento de Fisica Te\'orica and IFIC, Centro Mixto Universidad
de Valencia-CSIC, Facultad de Fisica, Burjassot-46100, Valencia, Spain}
\\[.1in]
Voja Radovanovi\'c
\footnote{E-mail: rvoja@rudjer.ff.bg.ac.yu} \\
{\it Faculty of Physics, P.O. Box 368, 11001 Belgrade, Yugoslavia}}

\def\AMV{Aleksandar Mikovi\'c
\footnote{On leave from Institute of Physics, P.O. Box 57, 11001 
Belgrade, Yugoslavia}
\footnote{Supported by Comissi\'on Interministerial de Ciencia y Tecnologia}
\footnote{E-mail: mikovic@lie1.ific.uv.es}
\\
{\it Departamento de Fisica Te\'orica and IFIC, Centro Mixto Universidad
de Valencia-CSIC, Facultad de Fisica, Burjassot-46100, Valencia, Spain}}

\def\endtitle{\par\end{quotation}\vskip3.5in minus2.3in\newpage}
 
 
\def\endabstract{\par\end{quotation}
        \renewcommand{\baselinestretch}{1.2}\small\normalsize}
 
 
\def\xpar{\par}                                         

\def\letterhead{
        \centerline{\large\sf INSTITUTE OF PHYSICS}
        \centerline{\sf P.O.Box 57, 11001 Belgrade, Yugoslavia}
        \rightline{\scriptsize\sf Dr Aleksandar Mikovi\'c}
        \vskip-.07in
        \rightline{\scriptsize\sf Tel: 11 615 575}
        \vskip-.07in
        \rightline{\scriptsize\sf E-mail: MIKOVIC@CASTOR.PHY.BG.AC.YU}}

\def\sig#1{{\leftskip=3.75in\parindent=0in\goodbreak\bigskip{Sincerely yours,}
\nobreak\vskip .7in{#1}\par}}

\def\ssig#1{{\leftskip=3.75in\parindent=0in\goodbreak\bigskip{}
\nobreak\vskip .7in{#1}\par}}

 
\def\ree#1#2#3{
        \hfuzz=35pt\hsize=5.5in\textwidth=5.5in
        \begin{document}
        \ttraggedright
        \par
        \noindent Referee report on Manuscript \##1\\
        Title: #2\\
        Authors: #3}
 
 
\def\start#1{\pagestyle{myheadings}\begin{document}\thispagestyle{myheadings}
        \setcounter{page}{#1}}
 
 
\catcode`@=11
 
\def\ps@myheadings{\def\@oddhead{\hbox{}\footnotesize\bf\rightmark \hfil
        \thepage}\def\@oddfoot{}\def\@evenhead{\footnotesize\bf
        \thepage\hfil\leftmark\hbox{}}\def\@evenfoot{}
        \def\sectionmark##1{}\def\subsectionmark##1{}
        \topmargin=-.35in\headheight=.17in\headsep=.35in}
\def\ps@acidheadings{\def\@oddhead{\hbox{}\rightmark\hbox{}}
        \def\@oddfoot{\rm\hfil\thepage\hfil}
        \def\@evenhead{\hbox{}\leftmark\hbox{}}\let\@evenfoot\@oddfoot
        \def\sectionmark##1{}\def\subsectionmark##1{}
        \topmargin=-.35in\headheight=.17in\headsep=.35in}
 
\catcode`@=12
 
\def\sect#1{\bigskip\medskip\goodbreak\noindent{\large\bf{#1}}\par\nobreak
        \medskip\markright{#1}}
\def\chsc#1#2{\phantom m\vskip.5in\noindent{\LARGE\bf{#1}}\par\vskip.75in
        \noindent{\large\bf{#2}}\par\medskip\markboth{#1}{#2}}
\def\Chsc#1#2#3#4{\phantom m\vskip.5in\noindent\halign{\LARGE\bf##&
        \LARGE\bf##\hfil\cr{#1}&{#2}\cr\noalign{\vskip8pt}&{#3}\cr}\par\vskip
        .75in\noindent{\large\bf{#4}}\par\medskip\markboth{{#1}{#2}{#3}}{#4}}
\def\chap#1{\phantom m\vskip.5in\noindent{\LARGE\bf{#1}}\par\vskip.75in
        \markboth{#1}{#1}}
\def\refs{\bigskip\medskip\goodbreak\noindent{\large\bf{REFERENCES}}\par
        \nobreak\bigskip\markboth{REFERENCES}{REFERENCES}
        \frenchspacing \parskip=0pt \renewcommand{\baselinestretch}{1}\small}
\def\unrefs{\normalsize \nonfrenchspacing \parskip=medskipamount}
\def\Item{\par\hang\textindent}
\def\Itemitem{\par\indent \hangindent2\parindent \textindent}
\def\makelabel#1{\hfil #1}
\def\topic{\par\noindent \hangafter1 \hangindent20pt}
\def\Topic{\par\noindent \hangafter1 \hangindent60pt}

\title{Quantum corrections for (anti)-evaporating  black hole}
{\MBVR}{}{July 2000}

\noindent  In this
paper we analyse the quantum correction for \Sch black hole in the
Unruh state in the framework of spherically symmetric gravity (SSG)
model. SSG is two-dimensional dilaton model which is obtained by
spherically symmetric reduction from four-dimensional theory. We find
the one-loop geometry of the (anti)-evaporating black hole and corrections for mass,
entropy and apparent horizon.

\endtitle

\section{Introduction}

Two-dimensional spherically symmetric gravity model (SSG) is 
interesting for many
reasons. This model is obtained from four dimensional (4D)
 Einstein-Hilbert action coupled
minimally to scalar fields by spherically symmetric reduction of
 metric and scalar fields. The reduction is done in the spirit of string
theory, via the introduction of  dilaton field $\Phi$,
assuming that the line element is of the form:
\be ds_{(4)}^2=g_{\mu\nu}dx^\mu dx^\nu +e^{-2\Phi}(\sin ^2\theta
d\f^2+d\theta^2)\ ,\label{reduc}\ee 
where $\mu , \nu =0,1$.
The action for this model is given by the equation (\ref{eq:S2})  below, and it has  \Sch black hole as a static vacuum solution.

One reason which makes this model interesting is that the 
quantum effective action for scalar fields
 can be calculated to the one-loop
order. This gives the possibility to obtain the backreaction effects of
quantized matter
 to  gravity analytically (in the case of black hole solution
 this is the backreaction of the Hawking radiation).
These analytic 2D calculations can then be compared with 
the numerical 4D estimates,
as the effective action can not be obtained analytically in 4D.
This analysis was done  in many details
for the \HH vacuum state of matter \cite{brm,bf,bf1}.
Summarizing, one can say that the main drawback
of the SSG model is that it gives the negative luminosity of the black
hole. It is  argued in the literature \cite{mwz} that this result is 
a consequence of the fact that only the radial
modes of the scalar field are counted in the 
expectation value of the energy density while
the angular modes are omitted. Formally, the negative luminosity 
 is not a surprising result as
the scalar field and the dilaton 
are strongly coupled at spatial infinity, as can be seen from the action
 (\ref{eq:S2}). There are 
also some
attempts to improve the lagrangian of the model \cite{kv99,klv98,klv99,lmr}.

2D dilaton gravity is also interesting by itself from the heuristic point
of view. Dilaton couplings are present in all theories 
which are obtained by
dimensional reduction from string theories.
Furthermore, the one-loop effective actions are nonlocal.
 One possibility to deal
with such  actions is their conversion to the local form by
introduction of auxiliary fields. 
The local form of action is rather handy
for calculations (e.g., for  equations of motion
or energy-momentum tensor). On
the other hand, the fact that auxiliary fields
describe nonlocal effects implies that they 
are dynamical, and it is a priori
unclear how to fix the arbitrary constants (or functions) in the
solutions. It is also not known whether all solutions have the physical
meaning. In the case of SSG model the properties of auxiliary fields are
rather well established for the \HH vacuum state. In the present paper
we extend the analysis to the Unruh vacuum. We think that it is of
importance to understand the ways  to describe
nonlocal effects by auxiliary fields. SSG is important as it 
provides us with an example of the effective
action which is tractable, but, as we shall see, in some respects
more complicated than the (usually discussed) Polyakov-Liouville
action.

The complementary way of discussing different vacuum states was
developed in the very instructive paper \cite{bf}  by Balbinot and Fabbri. 
Their analysis is based
on the conformal properties of fields under the change of the conformal vacuum state. In this method, the
initial step is to identify the energy-momentum tensor (EMT) of one
vacuum state (e.g., Boulware).
Then one can find the expectation values
of EMT in other states from conformal transformation properties
of fields.

The organization of the paper is the following. In section 2
we solve the equations 
of motion for the auxiliary fields in the Unruh vacuum
and obtain the value of energy-momentum tensor.
In order to fix the arbitrary functions in the solution we use the
conditions of regularity of EMT on the future horizon. For
comparison, the energy momentum tensor is found by \BF procedure.
The differences between \PoL action and SSG action are also discussed.
 In section 3 we find the influence of the Hawking radiation to the geometry
 in the one-loop order. In order to fix the integration constants
 in the metric, we impose the condition that the emitted flux
 of radiation is constant. We  calculate
  the ADM mass of the black hole. In
section 4 we obtain the position of the apparent horizon and 
entropy. Furthermore, we  analyse the behaviour of the entropy
along the line of the apparent horizon and find that the second law
of thermodynamics is fulfilled.

\section{Energy-momentum tensor and auxiliary fields}

The Einstein-Hilbert action with
minimally coupled $N$ scalar fields, $f_i$ ($i=1,\dots N$) in 4D is given by
\be \label{eq:S4}
\G_0^{(4)}={1\over 16\p G}\int d^4x\sqrt{-g^{(4)}}R^{(4)}-{1\over
8\pi }\sum _i\int d^4x\sqrt{-g^{(4)}}(\nabla f_i)^2\ \ .\ee
After spherically symmetric reduction (\ref{reduc}),
from the action (\ref{eq:S4})
we get two-dimensional classical action $\G _0$
\bea \G_0&=&{1\over 4G}\int d^2x\sqrt{-g}
\Bigl( e^{-2\Phi}(R+2(\nabla\Phi)^2+2e^{2\Phi})
-2Ge^{-2\Phi}\sum _i(\nabla f_i)^2 \Bigr)  \ ,\label{eq:S2} \eea
where $g$ and $R$ denote two-dimensional metric and curvature.
The Schwarzschild black hole is the classical vacuum solution 
of the equations of motion which follow from the action (\ref{eq:S2}). 
This solution is given by
\bea\label{eq:cs}
ds^2&=&-f(x^1)(dx^0)^2+{1\over f(x^1)}(dx^1)^2\ncr
 \F&=&-\log x^1\ncr
 f_i&=&0\ \ {\rm ( except\ at\ the\ point\ }x^1=0)\ , \eea
where $f(x^1)=1-a/x^1\ \ .$ The constant $a$ is the radius of the
event horizon, $a=2MG$, and $M$ is the
mass of the Schwarzschild black hole.

When we add the one-loop quantum correction for the matter fields $f_i$
to the  classical action (\ref{eq:S2}), we get
the nonlocal effective action. Its
one-loop part is given by \cite{bh,mr4,klv,no99,nooo99,mr3}:
\be \label{eq:S2eff} \bar\G_1=
-{N\over 96\p}\int d^2x\sqrt{-g}\Bigl( R{1\over \Box}R-12 R{1\over\Box }(\nabla\Phi )^2
+12R\Phi \Bigr)\ ,\ee
which describes the quantum effects of the scalar matter fields.
Calculations can be simplified  if the nonlocal 
correction part $\bar\G_1$
 is rewritten in the local form using two
auxilliary fields $\psi$ and $\chi$ \cite{brm}:
\be\label{eq:G1loc}
\G_1=-{N\over 96\p}\int d^2x\sqrt {-g} \left[2R(\psi -6\chi )+(\nabla
\psi)^2-12(\nabla \psi )(\nabla \chi)-12\psi (\nabla \F )^2+12R\F \right].
\ee
The additional fields $\psi$ and $\chi$ satisfy the equations of motion
\be\Box\psi =   R\ ,\label{11}\ee
\be\Box\chi =(\nabla \F)^2\ .\label{22}\ee
$\G _1$ and $\bar\G_1$ are equivalent in the following sense.
If we introduce the equations (\ref{11}-\ref{22})  into the local form 
of the action $\G _1$,
 we will get the nonlocal action $\bar\G_1$ up to boundary terms
\footnote{We would like to thank D. Vassilevich for the discussion 
considering this
point.}.
This difference does not influence the equations of motion. 
The analysis of the boundary terms can be postponed till the calculation
 of ADM mass and it was done carefully \cite{br00}.

The form of the action we will use is:
\bea \label{eq:S}\G=\G_0+ \G _1&=&{1\over 4G}\int d^2x\sqrt{-g}\Bigl( r^2R+2(\nabla
r)^2+2\Bigr) \ncr
&-& {\kappa\over 4G}
\int d^2x\sqrt{-g}\Bigl[ (\nabla\psi )^2+2R\psi -12(\nabla\psi )
(\nabla\chi ) \ncr
&-&12\psi{(\nabla r)^2\over r^2}-12R\chi -12R\log{r}
\Bigr)\Bigr], \eea
where $\k=NG\hbar /24\p$. Instead of the dilaton $\Phi$ we introduced
new variable, $r=e^{-\Phi}$.
Varying the action (\ref{eq:S}) we obtain the 
equations of motion \cite{brm}:
\be {\Box\psi =R}\ , \label{eqpsi}\ee
\be {\Box\chi ={(\nabla r)^2\over r^2} }\ , \label{eqchi}\ee
\be{ 2\Box r-rR= -6 \kappa \Big( 2\psi {\Box r\over r^2}
+2{(\nabla\psi )(\nabla r)\over r^2}-2\psi {(\nabla r)^2\over r^3}+
{R\over r}\Big) }\ , \label{eqr}\ee
\bea &&g_{\mu\nu}\big( \Box r^2-(\nabla r)^2-1\big) -2r\nabla _\mu\nabla _\nu r=
2G  T_{\mu\nu}=\ncr
&=&\kappa \Bigl( g_{\mu\nu}(2R+6\psi {(\nabla r)^2\over r^2}-{1\over 2}
(\nabla\psi )^2+
6(\nabla\psi )(\nabla\chi )-12{\Box r\over r}) \ncr
&+&\nabla _\mu\psi\nabla _\nu\psi -12\nabla _\mu\psi\nabla _\nu\chi
-2\nabla _\mu\nabla _\nu\psi +12\nabla _\mu\nabla _\nu\chi  \ncr
&+&12{\nabla _\mu\nabla _\nu r\over r}
-12(1+\psi ){\nabla _\mu r\nabla _\nu r\over r^2} \Bigr)\ . \label{eqg} \eea

First, let us note that $r=x^1$ ($\F =-\log x^1$) remains to be
the solution of the quantum-corrected equations of
motion (\ref{eqpsi}-\ref{eqg}), so we see that the  field $r$ has the
meaning of radius.
We will use the following notation for the coordinates: $x^1=r$, $x^0=t$.

We want to find the quantum correction of the geometry of 2D black hole 
for the case when the  black hole evaporates. This means that black hole is
in the Unruh state. Our calculation is perturbative in the
orders of  $\k$ which is 
a small parameter. All quantities will be calculated to the first order 
in $\k$, as the effective action is also calculated to this precision only. 
The ansatz for the one-loop metric is 
\be \label {eq:qs}
ds^2=
-F(r,\tilde v)e^{2\k \varphi}d\tilde v^2+
2e^{\k
\varphi}d\tilde vdr\ ,
\ee
and we  solve the equations in Eddington-Finkelstein $r$, $\tilde v$ 
coordinates,
\be \tilde v=t+r_*=t+r+a\log \,
({r\over a}-1)\ .\ee
The function $F$ is taken in the form
\be F(r,\tilde v)=f(r)+{\k m(r,\tilde v)\over r}=
1-{a\over r}+{\k m(r,\tilde v)\over r}\ .\ee
Introducing the ansatz (\ref{eq:qs}) into  (\ref{eqr}-\ref{eqg}),
we get that the equations for unknown functions $m$ 
and $\varphi$ in the first order in $\kappa$ take the simple form:
\be \k \pa_r\varphi=G{T_{rr}\over r}\ee
\be \k \pa_rm=2Ge^{-\k \varphi}T_{r\tilde v}\ee
\be \k \pa _{\tilde v}m=-2G(FT_{r\tilde v}+e^{-\k \varphi}T_{\tilde
v\tilde v})\ ,\ee
where $T_{rr}, T_{r\tilde v},{\rm and}\ T_{\tilde v\tilde v}$ are the
corresponding components of  the energy-momentum tensor defined by
the equation (\ref{eqg}).  The EMT is a quantity of the first
order in $\k$, so in order to determine it with the necessary precision 
 we need the zero-th order solution for 
 metric and
auxiliary fields. 

Let us briefly review how the solutions were found previously, in
\cite{brm}.
In the \HH state $\psi$ and $\chi $ are 
time-independent, as they describe the black hole in thermal
equilibrium with the Hawking radiation. Therefore, the solutions
 of the equations
(\ref{eqpsi}-\ref{eqchi}) are 
\be\psi = Cr +Ca\log{
r-a\over a}- \log {r-a \over r}\quad ,\ee 
\be\chi ^\prime ={2Dr^2-2r+a\over
2r(r-a)}\quad .\ee 
The assumption of regularity of EMT
on the classical horizon $r=a$ 
in the free-falling frame
gives the values of the integration constants:
$C={1\over a}$, $D={1\over 2a}$.

We will now solve the equations (\ref{eqpsi}-\ref{eqchi}) in the
general case. As mentioned, we need the zero-th order metric:
\be ds^2=g_{\mu\nu}dx^\m dx^\n=-f
d\tilde v^2+2d\tilde vdr\ .\ee
The other quantities entering equations  (\ref{eqpsi}-\ref{eqchi}) are
\be R=-{d^2f\over dr^2}\ ,\ \ {(\nabla r)^2 \over
r^2}={f\over r^2}\ .\ee
Introducing these values, the equation for $\psi$ becomes
\be \Box\psi =\partial _r(2\partial _{\tilde v}\psi +f\partial
_r\psi )=-{d ^2f\over d r^2}\ ,\ee
and it reduces to the linear partial differential equation:
\be 2\partial _{\tilde v}\psi +f\partial _r\psi
=-{d f\over d r}+\tilde\cg(\tilde v)\ .\label{psi1}\ee
In order to find the general solution of the equation (\ref{psi1})
one has to find two independent integrals
$\alpha (\tilde v, r, \psi
)={\rm const}$ and  $\beta (\tilde v,r,\psi )={\rm const}$ 
of the system
\be {d\tilde v\over 2}={dr\over f}={d\psi \over \tilde \cg(\tilde v)-\partial
_rf}\ \ ;\ee 
the general solution of (\ref{psi1}) is then an arbitrary function of
$\alpha$ and $\beta$. In our case, the independent integrals are
\be \alpha =r_*-{\tilde v\over 2}\ ,\ \ \beta =\psi +\log f -{1\over 2}\int
\tilde \cg(\tilde v)\, d\tilde v\ \ .\ee
Therefore, the general solution for $\psi$ can be written in the form
\be \label{eq:gspsi}\psi =-\log \left(1-{a\over r}\right)+ \cg(\tilde v)
+\cc (r_* -{\tilde v\over 2}) \ ,\ee 
where $r_*=r+a\log\left({r\over a}
-1\right)$, while $\cg(\tilde v)={1\over 2}\int \tilde \cg(\tilde v)d\tilde v$
and $\cc(r_*-{\tilde v\over 2})$ are arbitrary functions.    Similarly, the
equation for $\chi$  
\be \Box\chi =\partial _r(2\partial _{\tilde v}\chi +f\partial
_r\chi )={f\over \ r^2}\ ,\ee 
reduces to the system
\be {d\tilde v\over 2}={dr\over f}={d\chi \over \tilde \ch(\tilde v)
+{a-2r\over 2r^2}}\ .\ee
The general solution for $\chi$ is  
\be \label{eq:gschi}\chi=-{1\over 2}\log{r(r-a)\over a^2} +\ch(\tilde v)+ \cd(r_*
-{\tilde v\over 2})\ ,\ee 
where $\ch(\tilde v)$ and $\cd(r_* -{\tilde v\over
2})$ are arbitrary functions. The functions $\cg(\tilde v), \ 
\cc(r_*-{\tilde v\over 2}),\ \ch(\tilde v)$ and $\cd(r_* -{\tilde v\over 2})$
describe various quantum states of  matter.   To recover the static \HH vacuum
solution we have to put all functions linear in their arguments in 
order to cancel $t$-terms. This, combined with the condition of
regularity  gives  $\cc(r_*
-{\tilde v\over 2})= {1\over a}(r_* -{\tilde v\over 2}),\ \cg(\tilde v)={1\over
2a}\tilde v, \ \ch(\tilde v)={1\over 4a}\tilde v$ and $\cd(r_* -{\tilde v\over
2})={1\over 2a}(r_* -{\tilde v\over 2}). $ 

We now pass to the case of the Unruh vacuum.
It  is most naturally discussed in the
null-coordinates $u$, $v$:
\be v=\tilde v\ ,
 u=\tilde v-2r_*=\tilde v-2\left(r+a\log({r\over a}-1)\right)\ .\ee

The Unruh vacuum state is  defined as the state which has the 
EMT regular on
the future event horizon, $u\to \infty ,\ v=$ constant. 
The conditions of  regularity in the free falling frame read \cite{cf}:
\be T_{vv}<\infty\ ,
\ \  {T_{uv}\over f}<\infty \ , 
\ \  {T_{uu}\over f^2}<\infty \ .\ee
Components of the energy-momentum tensor in the $u,\ v$ coordinates can
be found from the relations
\be T_{rr}=4\left({r\over r-a}\right)^2T_{uu}\ee
\be T_{r\tilde v}=-2{r\over r-a}(T_{uu}+T_{uv})\ee
\be T_{\tilde v\tilde v}=T_{uu}+2T_{uv}+T_{vv}\ee
Along with the condition of regularity of EMT, we will impose that
at the spatial infinity $r\to\infty$ the outgoing flux $T_{uu}$ 
has a constant nonvanishing value, while the ingoing flux
$T_{vv}$ tends to 0. When we introduce the solutions 
(\ref{eq:gspsi}), (\ref{eq:gschi}) for the components of EMT we get:
\be T_{uv}={a\over 24\pi}{r-a\over r^4}\ee
\bea T_{vv}&=&{(a-r)^2\over 16\pi r^4}\log{r-a\over r}+{1\over
48\pi}(\cg ^{\prime 2}-12 \cg ^\prime \ch ^\prime -2\cg ^{\prime \prime}+12
\ch ^{\prime \prime})\ncr
&-&{1\over 192\pi r^4}\Bigl( -3a^2+4ar+12(a-r)^2\cc +12(a-r)^2\cg \ncr
&+&(12ar^2-24r^3)\cg ^\prime\Bigr) \label{TVV}\eea
\bea T_{uu}&=&{(a-r)^2\over 16\pi r^4}\log{r-a\over r}+{1\over
48\pi}(\cc ^{\prime 2}-12 \cc ^\prime \cd ^\prime -2\cc ^{\prime \prime}+12
\cd ^{\prime \prime})\ncr
&-&{1\over 192\pi r^4}\Bigl( -3a^2+4ar+12(a-r)^2\cc +12(a-r)^2\cg \ncr
&+&(6ar^2-12r^3)\cc ^\prime\Bigr) \label{TUU}\eea
(primes denote derivatives of the functions 
with respect to their arguments).

There is no information about the unknown functions 
contained in $T_{uv}$. Further, it can be seen that
$T_{vv}$ is regular on the horizon. The condition that $T_{vv}\to 0$
as $r\to\infty$ means that in this limit
\be \cg ^{\prime 2}-12 \cg ^\prime \ch ^\prime -2\cg ^{\prime \prime}+12
\ch ^{\prime \prime} =0\ .\ee
The solution of the last equation, which is in accordance with the 
radiation law, is given by  linear functions
$\cg$, $\ch$:
\be  \cg (\tilde v)=g\,\tilde v \ ,\ \ \ch (\tilde v)=h\,\tilde v \ ,\ee
with
\be g(g-12h)=0\ ,\label{gh}\ee i.e. either $g=0$ or $g=12h$.

Similarly, the condition that  $T_{uu}\to$ const
as $r\to\infty$ gives that the functions $\cc$ and $\cd$ are linear in their arguments,
\be \cc (x)=c\, x \ , \ \ \cd (x)=d\, x \ .\ee
Nonsingularity of ${T_{uu}\over f^2}$ on the horizon
gives us the values of the constants: 
$c={1\over a}$, $d={1\over 2a}$. 
Introducing $c$ and $d$ in (\ref{TUU}) we see that the luminosity 
has the \HH value  $-{5\over 192\pi a^2}$. 2D black hole antievaporates.
 This is because we took into account the contribution of the 
 $s-$modes of the radiation only.  

To conclude our reasoning, let us observe that one arbitrariness  remained, and that is the 
dependence of EMT on the constant $g$. This arbitrariness can be naturally
fixed by choosing the $g=0$ solution of the condition (\ref{gh}).
Note also that the value of the constant $h$ does not enter EMT, and
therefore we can fix it freely, e.g.  $h={1\over 4}$. 
Finally we have the solution for $\psi$, $\chi$ in the zero-th order 
\be \psi = {r\over a}+\log {r\over a}-{v\over 2a}\ ,\ee
\be \chi ={r\over 2a}-{1\over 2}\log {r\over a}\ .\ee
We just mention briefly that it can be shown that for $g=0$
the value of $h$ does not influence the ADM mass.

We can now perform the
Balbinot-Fabbri procedure \cite{bf}
and compare the values of EMT.
If the vacuum state of
matter is defined in such a way 
 that the ingoing and outgoing modes have positive frequency
with respect to the coordinates $u,v$, the EMT
corresponds to the Boulware state:
\be \bra{u,v}\,\hat T_{uv}\,\ket{u,v}=-{1\over 12\p}
(\partial _v\partial _u\rho +3\partial _v\Phi\partial _u\Phi
-3\partial _v\partial _u\Phi )
\ ,\ee
\bea \bra{u,v}\,\hat
T_{vv}\,\ket{u,v}&=&-{1\over 12\p}
(\pa _v\rho\pa _v\rho-\pa ^2_v\rho)+{1\over 2\p}
\Big( \rho(\partial _v \Phi )^2+{1\over 2}{\partial_v\over\partial _u}
(\partial _v\Phi \partial _u\Phi )\Big) \ncr
&-&{1\over 4\p}( -2(\partial _v\rho)(\partial _v\Phi )+
\partial^2 _v\Phi )\ ,\ncr\eea
\bea \bra{u,v}\,\hat
T_{uu}\,\ket{u,v}&=&-{1\over 12\p}
(\pa _u\rho\pa _u\rho-\pa ^2_u\rho)
+{1\over 2\p}
\Big( \rho(\partial _u \Phi )^2+{1\over 2}{\partial _u\over\partial _v}
(\partial _+\Phi \partial _-\Phi )\Big) \ncr
&-&{1\over 4\p}( -2(\partial _u\rho)(\partial _u\Phi )+
\partial^2 _u\Phi )\ ,\ncr\eea
where $\rho ={1\over 2}\log (1-{a\over r})$ is a conformal factor.

The conformal transformation to the
other conformal state  $\sket{\tilde u, \tilde v}$ 
defined by the other set of null-coordinates 
$\tilde u=\tilde u(u)$, $\tilde v=\tilde v(v)$,
gives 
\be \bra{\tilde u, \tilde v}\,\hat T_{uv}\,\ket{\tilde u, \tilde v}
=\bra{u,v}\,\hat T_{uv}\,\ket{u,v}\ ,\ee
\bea \bra{\tilde u,\tilde v}\,\hat T_{vv}\,\ket{\tilde u,\tilde v}&=&
\bra{u,v}\,\hat T_{vv}\,\ket{u,v}
+{1\over 24\p}\left({G^{\prime\prime}\over G}
-{1\over 2}{G^{\prime 2}\over G^2}\right)\ncr
&+&{1\over 4\p}\left( (\partial _v\Phi )^2\log (FG)+{G^\prime\over G}
\int du \partial _v\Phi \partial _u\Phi   \right) \ , \label{eq:EMTbf}\ncr \eea
\bea \bra{\tilde u,\tilde v}\,\hat T_{uu}\,\ket{\tilde u,\tilde v}&=&
\bra{u,v}\,\hat T_{uu}\,\ket{u,v}
+{1\over 24\p}({F^{\prime\prime}\over F}
-{1\over 2}{F^{\prime 2}\over F^2})\ncr
&+&{1\over 4\p}\Big( (\partial _u\Phi )^2\log (FG)+{F^\prime\over F}
\int dv (\partial _u\Phi )(\partial _u\Phi )  \Big) \ ,
 \label{eq:EMTbf1}\ncr \eea
where $F(u)={du \over d\tilde u}$, $G(v)={dv\over d\tilde v}$.

Unruh vacuum state is the state $\sket{U,v}$, $U$ being
the Kruskal coordinate
$U=-2ae^{{u\over 2a}}$.
 Using (\ref{eq:EMTbf}-\ref{eq:EMTbf1}) after  simple
calculation, we get the value of the EMT in the Unruh state
 (${1\over 24\pi}={\k\over G}$) :
\be \label{tuv}
T_{uv}={\k\over G} (1-{a\over r}){a\over r^3}\ee
\bea \label{tuu}
T_{uu}&=&{\k\over G} \Big({3a^2-4ar\over 8r^4}-{5\over 8a^2}-{3\over 2a}({a\over
2r^2}
-{1\over r})\ncr
&+&{3\over 2r^2}(1-{a\over r})^2({v\over 2a}-{r\over a}-\log{r\over a})\Big)
\ \ncr\eea
\be \label{tvv}
T_{vv}={\k\over G} \Big({3a^2-4ar\over 8r^4}
+{3\over 2r^2}(1-{a\over r})^2({v\over 2a}-{r\over a}-\log{r\over a})\Big)
\ .\ee

These expressions are the same as the previously given
(\ref{TVV}-\ref{TUU}) with fixed integration functions.

Let us give one final comment of the values of EMT (\ref{tuv}-\ref{tuu}).
The obtained values have $v$-dependenence, i.e.
$t$-dependence. This dependence does not show up in the asymptotic
behaviour of EMT and it was considered by \cite{bfs} as an unwished
property of the energy-momentum tensor. 
In fact, in \cite{bfs} the auxiliary fields were constrained in such a way
that the time-dependence of $\psi$, $\chi$ would not produce any time
dependence in EMT. We think that a condition 
like this is too stringent and unneccessary. It
holds, though, in the "minimal coupling" case, i.e. in the case when
the effective action is given by the \PoL term only, as it can easily be seen.
Namely, it is known \cite{bf1} that in this case
the change of the conformal frame 
produces in EMT only the additional term proportional to the Schwarzian derivative
of the transformation of coordinates:
\be \bra{\tilde u,\tilde v}\,\hat T_{vv}\,\ket{\tilde u,\tilde v}=
\bra{u,v}\,\hat T_{vv}\,\ket{u,v}
+{1\over 24\p}\Big({F^{\prime\prime}\over F}
-{1\over 2}{F^{\prime 2}\over F^2}\Big)\ .\ee
For exponential mappings, which are typical for the
transformation to Kruskal coordinates, the Schwarzian derivative is
constant. This means that if we start with the time-independent
EMT for, e.g., \HH vacuum, we will get the time-independent 
EMT for all other conformal vacua. But this is the special property
 of the \PoL
effective action.
In SSG case the structure of the additional terms is more complicated
and this brings the time-dependence in 
the Unruh vacuum state. The
fact that this dependence is linear is in accordance with the
expected property that the black hole in the Unruh vacuum radiates at
constant rate,  ${dM(t)\over dt}=$ const. The meaning of the mass
 $M(t)$ will be discussed in more details after we solve the
backreaction equations for the metric and identify the ADM mass of
the solution.

\section{Backreaction and corrected geometry}

The equations which determine the one-loop correction of the metric
can now easily be integrated. The solution is:
\be
\varphi={5\over ar}+3{a-2 v\over 4ar^2}
+{3\over r^2}\log{r\over a}-{5\over 2a^2}\log{r\over l}+C_1\ee
\bea
m&=&{5r\over 2a^2}+{a+6 v\over 2ar}+{11a-6v\over 4r^2}-{5
v\over 4a^2}\ncr
&-&3{2r-a\over r^2}\log{r\over a}+{5\over 2a}\log{r\over l}+C_2 \ncr\eea
We see that the functions $m( v, r)$ and $\varphi ( v, r)$ 
depend linearly of $ v$, i.e. of
time. There are two independent integration constants, $C_1$ and  $C_2$.  
The expression for the ADM energy  was
found in \cite{br00}. The value of the energy is given by the value of the
boundary term which has to be added to the canonical hamiltonian in
order to have a well defined theory. It is given by
\be \Delta =-\delta H_b\ ,\ee
where
\bea  \label{delta} 4G\D
&=& {\sg\over\jj} (4B\rp \delta r-2\kappa\psip\delta\psi
+12\kappa\psip\delta\chi +12\kappa\chip\delta\psi )\ncr
&+& {2\over\sg}\delta ({-g\over\jj})(Ar\rp -\kappa\psip +6\kappa\chip )+
{2\over\sg}\Big({-g\over\jj}\Big)^\prime (Ar\delta r -\kappa\delta\psi +
6\kappa\delta\chi )\ncr
&+&4G\pi ^{11}(2\delta\nj -\raz\delta\jj )+
4G\raz(\pi _r\delta r+\pi _\psi
\delta\psi +
\pi _\chi \delta\chi )\ .\eea
 $\delta$ denotes the variation in the 
chosen class of field configurations, described in more details in \cite{br00}.
 $A$ and $B$ are  $A= 1+{6\k\over r^2}\ , B=1+{6\k\psi\over r^2}\ .$  Of course, in order to identify the real value of energy,
we have to find it in a coordinate system which is asymptotically Minkowskian.
As we have solved the equations for $m$ and  $\varphi$, we can now write the
corrected values of the components of metric: \bea
g_{00}&=&-\Big(1-{a\over r}+{\k m\over r}+2\k (1-{a\over r})\varphi\Big)\ncr
g_{01}&=&-\kappa \Big({m\over r-a}+\varphi \Big) \ncr g_{11}&=&{r\over r-a}
-\kappa {mr\over (r-a)^2} \ ,\label{123}\eea 
so we see that, unlike the static case,
the metric is not diagonal in the first order in $\kappa$.

In order to find a coordinate system 
$\tilde t,\ \tilde r$ in which the asymptotic values
of the metric are 
\be \tilde g_{00}\to -1+O({1\over L})
\ ,\ \ \tilde g_{01}\to 0\label{asy}\ee
(it is not really necessary to assume also $\tilde g_{11}\to 1$, as we are
interested only in the value of the energy),
we introduce the transformation of coordinates
\be \tilde t=t+\kappa\alpha (t,r)\ ,\ \ \tilde r =r\ .\ee
Under this transformation, the metric transforms as
\bea \tilde g_{00}&=&g_{00} (1-2\kappa {\partial\alpha\over\partial t})\ncr
\tilde g_{01}&=&g_{01}- \kappa {\partial\alpha\over\partial r}g_{00}\ncr
\tilde g_{11}&=&g_{11} \ .\eea
In accordance with the
asymptotic  relations 
(\ref{asy}) the function $\a$ should be choosen in the
form
\be \a (t,r)= F_1r+F_2t+F_3rt+F_4t^2\ ,\ee
where
\be F_1={5\over 4a^2}+{9\over aL}-{5\over 2a^2}\log {L\over l} -{5\over 4aL}
\log ({L\over a}-1)+{LC_1\over L-a}+{LC_2\over (L-a)^2}\ee
\be F_2=
{15\over 8a^2}+{45\over 8aL}-{5\over 2a^2}\log{L\over l} -{5\over 8aL}
\log {L\over a}+C_1+{C_2\over 2(L-a)
}\ee
\be F_3=-{5\over 4a^2L}\ee
\be F_4={5\over 32a^2(L-a)}-{5\over 16a^2L}\ . \ee
 The coordinate transformation induces
the following change in the boundary term:
\be 4G\tilde\Delta =4G\D -2\k 
r\left({\pa \a\over \pa t}\d (-{g\over g_{11}})+2(-{g\over g_{11}})\d ({\pa
\a\over \pa t})\right) \ .\ee
Introducing the obtained solutions for $\psi$, $\chi$, $g$
we get for the value of $\tilde\D$:
\be 4G\tilde\D=-2\d a +\k\Big( {21\over 4a^2}-
{11L\over 2a^3}-{C_2(a)\over L-a}-{5\over a^2}\log{L\over l}+
{5\over a^3}t\Big)\d a     \ .\ee 
The corresponding value of energy is
\be \label {adm} \tilde H_b=-{1\over 4G}\int 4G\tilde\D =
M+{\k \over 4G}\Big({21\over 4a}-{11L\over 4a^2}-{5\over a}\log{L\over
l}+{5\over 2a^2}t+\int{C_2\over L-a}da\Big) \ee 
The first term in (\ref{adm}) is the classical mass of the black hole, while the
second one is the quantum correction of the mass. We can take that $C_2=0$. 
One immediately notes the time-dependence of ADM mass, which is in
agreement with radiation law of the black hole. Namely, 
\be {d\tilde H_b \over
dt}=T_{uu}\mid_{r\to L}=-{5\over 192\pi a^2}.\ee 
The increase of the mass 
cooresponds to the fact that the outgoing flux is
negative  at large distances, e.g. that the black hole antievaporates.
It is important to mention that the mass increases only if we consider
large but finite volumes $L$. If we take
the limit  $L\to \infty$, 
the $t$-term in the expression for energy (\ref{adm}) can be
neglected in comparison with the larger terms
proportional to $\log L$ and $L$, so we have the conservation
of the energy of the whole system, $\dot{ \tilde H_b}=0$. Notice, that  "mass
 function" $M(r,v)=M-\k{m(r,v)\over 2}$ satisfies the condition $\dot{M}(r,v)=-5/192\p a^2$.
\section{Apparent horizon and entropy}

Apparent horizon is the boundary of the trapped surfaces. In 2D dilaton gravity 
it is defined by \cite{ruso}
\be \label{trap}  g^{\mu\nu}\partial _\mu r\partial _\nu r =0\ .\ee
 If we define the one-loop corrected
null coordinates by 
\be ds^2= -e^{2\r} d\bar u\, d\bar v\ee
the condition (\ref{trap}) is reduced to $\pa _{\bar u}r=0$ and $\pa
_{\bar v}r=0\ .$
We will take $\bar v=v=t+r_*$. The other null coordinate  $\bar u$ can be 
found easily.
The first step is to rewrite the metric (\ref{eq:qs})  in
the form
  \bea
ds^2&=&-Fe^{2\k \varphi }(d\bar v-{2\over F}e^{-\k \varphi}dr)d\bar v
   \ncr
   &=&-{Fe^{2\k \varphi }\over \m}(\m d\bar v-{2\mu \over F}e^{-\k \varphi}
   dr)d\bar v\ ,
   \eea
where $\m $ is the integration factor. Therefore, the conformal coordinate
$\bar u$  satisfies
\be \label{eq:dut}
d\bar u=\m d\bar v -{2\m \over F}e^{-\k \varphi}dr\ .\ee
We will not solve the previous equation for $\bar u$, but just use it to find 
the position of the apparent horizon. From (\ref{eq:dut})
we get
\be dr={1\over 2}e^{\k \varphi}F(d\bar v-{1\over \m}d\bar u)\label{eq:dr}\ .\ee
The last equation, if we use
$\pa _{\bar u}r=0$ and $\pa_{\bar v}r=0\ $
implies
$e^{\k \varphi}F=0$ on the horizon. This means that 
 the equation of the apparent horizon is
\be \label{AH}  (1+\k \varphi )(1-{a\over r}+{\k m\over r})=0\ .\ee
The position of the apparent horizon is found  perturbatively
taking $r_{AH}=a+\k r_1$, where $r_1$ is the first-order correction. From
 equation (\ref{AH}) we get
\be \label{eq:ah}
r_{AH}=a-\k\Bigl({23\over 4a}+{5\over 2a}\log {a\over
l}+{1\over 4a^2}\bar v\Bigr)\ .\ee

The intersection point between the line of singularity and the
apparent horizon is the endpoint of the Hawking radiation. It is
given by
\be \bar u_{int}=\infty\ , \bar v_{int}=4a^2\Big({ a\over \k} -{23\over
4a}-{5\over 2a}\log {a\over l}\Big)\approx{4a^3\over \k}\ .
\ee
As we can take the $\bar v$-coordinate as
the time, we see that the (anti)-evaporation of the black hole is very long but finite.

In order to calculate the entropy of the quantum corrected 
solution, we  use the  Wald
technique \cite{w}.
Note that the conical singularity method is 
 defined for static configurations only and therefore cannot be used here.
 In references \cite{m,jkm,iw} it was shown
that for the lagrangians of the form $L=L(f_{m}, \nabla f_{m},
g_{\m\n} , R_{\m\n\r\s})$ ($f_{m}$ are the matter fields)
the entropy is given by
$$S=-{2\p}\e _{\a\b}\,\e_{\c\d}\,{\pa L\over \pa R_{\a\b\c\d}}\Big|_H \ ,$$
evaluated on the horizon.
In our case we find
\bea \label{ent}
S&=&{\p \over G}\Big(r^2-\k (2\psi -12\chi -12\log r)\Big)\Big|_{AH}\ncr
&=&{\p \over G}\Big(r^2+\k (4{r\over a}-8\log {r\over a}+12\log {r\over l}+{1\over
a}\bar v)\Big)\Big|_{AH}\ncr
&=&{\p\over G}\Big(a^2-\k({15\over 2}
+5\log{a\over l}-{\bar v\over 2a})\Big)\ .\ncr
\eea
Now, we will show that the entropy increases along the line of apparent horizon. This end
 we will find the equation for $\bar u$ coordinate. The integration factor, which we introduced in (\ref{eq:dut}), is of the form
\be\label{ans}
\m=1+\k R(r)+\k V(\bar v) \ ,\ee
where $R(r)$ and $V(\bar v)$ are unknown functions.
If we introduce the ansatz (\ref{ans}) in the condition of integrability of equation (\ref{eq:dut}),
\be {\pa \m \over \pa r}\Big|_{\bar v}=-2{\pa \over \pa \bar v}\Big({\m\over F}e^{-\k\varphi}\Big)\Big|_r\ ,\ee
we obtain
 the following expressions
\be \label{V}
V(\bar v)=\a \bar v\ ,\ee
\be \label {R}
R(r)=-2\a r-{1\over 2a(r-a)}-{4a^3\a +5\over 2a^2}\log(r-a)\ ,\ee
where $\a$ is the integration constant. On the other hand, if we start from
\be {\pa \bar u\over \pa \bar v}\Big|_r=1+\k R(r)+\k V(\bar v)\ ,\ee
\be {\pa \bar u\over \pa r}\Big|_{\bar v}= -{2\m\over F}e^{-\k \varphi}\ee
we get 
\be\label {baru} \bar u=\bar v +\k \bar vR(r) +{1\over 2}\k \a\bar v^2 +G(r)\ .\ee
Therefore the function $G(r)$ is determined by equation
\bea\label {G}
{dG\over dr}&=&-{2r\over r-a}\Big[1+\k \Big(-2\a r-{1\over 2a(r-a)}-{5+4\a a^3\over 2a^2}\log(r-a)\Big)\ncr
&-&\k\Big({5\over ar}+{3\over 4r^2}+{3\over r^2}\log{r\over a}-{5\over 2a^2}\log{r\over a}+C_1\Big)\ncr
&-&\Big({\k\over r-a}({5r\over 2a^2}+{1\over 2r}+{11a\over 4r^2}
-3{2r-a\over r^2}\log{r\over a}+{5\over 2a}\log{r\over l}\Big)\Big]\ ,\ncr\eea
which can easily be integrated. 
The derivative of the entropy along the apparent horizon is determined by
\be t^a\pa _aS=\Big({\pa \over \pa \bar v}+{d\bar u_{AH}\over d\bar v}{\pa \over \pa \bar u}\Big)S\ ,\ee
where  $t^a$ the is tangent vector of the apparent horizon. 
The expression (\ref{eq:ah}) for the apparent horizon and (\ref{V}-\ref{G})
give
\be t^a\pa _aS= {\k \p\over 2aG}>0\ .\ee
So, the entropy increases along the line of apparent horizon. 
This shows that the second law of thermodynamics is fulfilled in the
framework of the SSG model.

\section{Conclusions}

In this paper we calculated the backreaction effects of the Hawking radiation 
in the Unruh state of the \Sch black hole.
The effect is discussed in the framework of the SSG model.
 The calculation was simplified 
 using the formalism of  auxiliary fields. It is shown that the definition 
 of the Unruh state fixes the integration functions 
 and that the corresonding EMT coincides with EMT 
 calculated by other methods. The position of the apparent horizon is found 
 and the evaporation of the black hole discussed. 
 The obtained duration of the evaporation is large (proportional 
to $1/\kappa $). Unfortunately, at the
intersecting  point of the line of singularity and apparent horizon  
the singularity becomes  naked, which prevents us from predicting the
 future evolution of the black hole.
The discussion of the static remnant of the black hole is an interesting
question and will be the subject of futher investigation. The entropy of the black
 hole-radiation system is obtained and it is shown it increases during the evolution.
 The quantum corrections of the energy of the system
 are calculated using the ADM procedure.  
 We found that the flux of the radiation through the large spherical 
 surface of the radius $L$ is in accordance with the radiation law. 
 In the limit $L\to \infty$ though,
 the energy of the whole system is conserved, as one would expect.

{\bf Acknowledgments} The authors want to thank Albert Einstein
Institute f\" ur Gravitationsphysik in Potsdam, Germany, and 
Erwin Schr\"odinger Institute
in Wien, Austria for their hospitality, as the parts of this
paper were done during our stay in these institutes.

\end{document}